\documentclass[a4paper,11pt]{article}
\pdfoutput=1 

\usepackage{jcappub} 

\usepackage[T1]{fontenc} 

\title{\boldmath Dark Energy within the Generalized Uncertainty Principle in Light of DESI DR2}

\author[a,b,c,d,1]{A. Paliathanasis,\note{Corresponding author.}}

\affiliation[a]{Department of Mathematics, Faculty of Applied
Sciences,\\ Durban University of Technology, Durban 4000, South Africa}
\affiliation[b]{Centre for Space Research, North-West University,\\Potchefstroom 2520, South Africa}
\affiliation[c]{Departamento de Matem\`{a}ticas, Universidad Cat\`{o}lica del Norte,\\Avda.
Angamos 0610, Casilla 1280 Antofagasta, Chile}
\affiliation[d]{National Institute for Theoretical and Computational Sciences (NITheCS), South Africa}

\emailAdd{anpaliat@phys.uoa.gr}

\abstract{In this study, we modify the $\Lambda$CDM model by introducing a deformed
algebra within the framework of the Generalized Uncertainty Principle (GUP).
We formulate the modified Raychaudhuri equation, where new terms are
introduced which describe dynamical pressure components. For the quadratic GUP
model, we derive the Hubble function, which leads to a time-dependent dark
energy model. The free parameters are determined using late-time observational
data, the Pantheon+ SNIa sample, the cosmic chronometers, and the DESI 2025
BAO data. We find that the modified model introduce only one new additional
degree of freedom compared to the $\Lambda$CDM model. The GUP-Modified
$\Lambda$CDM model provides a better fit to the data than the undeformed
theory. {According to Jeffrey's scale for Bayesian evidence, we find
weak support in favor of the GUP-Modified model. }Furthermore, we compare the
same model with the DESI 2024 BAO data and find that the Bayesian evidence
becomes stronger with the inclusion of the DESI 2025 release.}

\begin{document}
\maketitle
\flushbottom

\section{Introduction}

\label{sec1}

The cosmological constant, which leads to the $\Lambda$CDM model, is the
simplest dark energy model, as it yields a set of integrable cosmological
field equations capable of explaining cosmic acceleration with the minimum
number of free parameters. However, the new release from the DESI\ 2025
collaboration \cite{des4,des5,des6} indicates a preference for dynamical dark
energy models over the $\Lambda$CDM model. The family of $w_{0}w_{a}$CDM
parametric models was examined, and it was found that they fit the
cosmological data better than the~$\Lambda$CDM model. This new release is
consistent with the conclusions of the DESI 2024 collaboration
\cite{des1,des2,des3}. Furthermore, a series of independent studies
\cite{ra1,ra2,ra3} analyzing various cosmological datasets---beyond those from
DESI---also indicate a preference for dynamical dark energy models.

From a phenomenological perspective, there exists a wide range of such models,
with the simplest being the Chevallier-Polarski-Linde (CPL) parametrization
\cite{cpl1,cpl1a}. Nevertheless, due to the limitations of the CPL model,
numerous alternative dynamical dark energy models have been proposed in the
literature
\cite{par1,par2,par4,par5,par6,par7,exp1,exp2,osc1,osc2,osc3,osc4,osc5,osc6,osc7}%
. However, the theoretical description of these models remains an open
problem. An important characteristic of parametric dark energy models is that
they allow for the possibility of phantom crossing for the dark energy
equation of state parameter.

In the literature, there are two main directions for investigating the nature
of dark energy. One approach introduces fluid components with negative
pressure into Einstein's General Relativity to drive the dynamics and account
for cosmic acceleration. Scalar fields, Chaplygin gases, and k-essence are
among the various fluid models proposed by cosmologists over the past decades
\cite{sf1,sf2,sf3,sf4,sf5,sf6,sf7,sf8,sf9,sf10,sf11}. An alternative approach
involves modifying Einstein's Hilbert action by introducing new geometric
invariants. This leads to the emergence of new degrees of freedom and
effective fluid components of geometric origin that can account for cosmic
acceleration. In this framework, dark energy has geometric origin which follow
from the gravitational theory
\cite{md1,md2,md3,md4,md5,md6,md7,md8,md9,md10,md11}.

{Quantum effects provide potential mechanisms to explain the phantom
behaviour of the dark energy. Indeed, in \cite{nf1} it was found that for two
quintessence scalar fields the equation of state parameter cross the phantom
line due to the existence of free or bound states for the Casimir operator of
the symmetry algebra of the fields. On the other hand, GUP is a potential
mechanism to explain dynamical dark energy in a }$\Lambda${CDM universe
\cite{nf2}. A similar ranalysis performed recently in \cite{nf3} by using the
extended uncertainty principle (EUP), where it was found that EUP can lead to
phantom-like dark energy.}

In this work, we follow a different approach by modifying the Friedmann
equations of the $\Lambda$CDM model using a deformed algebra within the
framework of the Generalized Uncertainty Principle (GUP) \cite{rev0,rev0a}. In
GUP, a minimum length emerges in Heisenberg's uncertainty principle. As a
result, a modified deformed algebra arises, which alters the physical laws
that the system must satisfy. GUP has found numerous applications in the study
of astrophysical objects \cite{bh1,bh2,bh3,bh4,bh5,bh6,bh7}, and various
studies have explored its cosmological implications
\cite{de1,de2,de3,de4,de4a,de4b,de4c,de4d}, including its relevance to the
dark energy problem \cite{de5,de6,de7,de8,de10} through the existence of a
minimum length. For a review on GUP, we refer the reader to \cite{rev1}.

The Friedmann equations for the $\Lambda$CDM model admit a Hamiltonian
formalism. To remain consistent with previous studies on the effects of GUP in
Szekeres spacetimes \cite{angup}, we express the Friedmann equations and the
corresponding Hamiltonian in terms of the energy density of cold dark matter
(CDM). By defining a new, modified Hamiltonian system incorporating the GUP
modifications \cite{hm1,hm2}, we derive a modified Raychaudhuri equation. In
this formulation, the additional terms introduced by the deformed algebra
transform the cosmological constant into a dynamical dark energy model. We
introduce dimensionless variables to express the field equations, which also
allows us to define dimensionless functions that describe the deformed algebra.

We consider a quadratic GUP parametrization. We explicitly derive the
asymptotic solution of the GUP-Modified cosmological equations. The quadratic
GUP results in a model, where the cosmological constant is described as a
dynamical dark energy fluid characterized by a single parameter. This model
provides a unified dark sector description, in which both dark matter and dark
energy are described from the same underlying function. We examine this
analytic solution model and constrain it using late-time cosmological data.
The structure of the paper is as follows.

In Section \ref{sec2}, we present the basic elements of the GUP and the
application of the deformed algebra to the modification of Hamilton's
equations. Section \ref{sec3} contains the main theoretical framework of this
study. In particular, we consider the $\Lambda$CDM model and apply the GUP to
modify the Raychaudhuri equation, following the procedure established in
\cite{angup}.

In Section \ref{sec4}, we focus on the quadratic GUP and study the asymptotic
dynamics of the modified cosmological equations. Moreover, we explicitly solve
the field equations and derive the modified Hubble function in terms of
closed-form expressions. In Section \ref{sec5}, we test the GUP-Modified
models using late-time cosmological data, including the Pantheon+ supernova
dataset (SNIa), cosmic chronometers (CC), and Baryon Acoustic Oscillations
(BAO) data from the DESI 2024 and DESI 2025 releases. The results are compared
with those of the $\Lambda$CDM model using the Akaike Information Criterion
and Jeffrey's scale for Bayesian evidence. Finally, in Section \ref{sec6}, we
present our conclusions.

\section{Generalized Uncertainty Principle}

\label{sec2}

We introduce the Generalized Heisenberg's Uncertainty Principle relation
\cite{Quesne2006,Vagenas,Kemph1,Kemph2,Vag1,Vag2}
\begin{equation}
\Delta X_{i}\Delta P_{j}\geqslant\frac{\hbar}{2}[\delta_{ij}(1+\beta
P^{2})+2\beta P_{i}P_{j}], \label{gp.01}%
\end{equation}
where $\beta=\frac{\beta_{0}}{M_{Pl}^{2}c^{2}}=\frac{\beta_{0}\ell_{Pl}^{2}%
}{2\hbar^{2}}$, $\beta_{0}$ is a deformation parameter and is of the scale
$O\left(  1\right)  $, $M_{Pl}$ is the Planck mass, $\ell_{Pl}$ is the Planck
length, and $M_{Pl}c^{2}$ is the Planck energy. Uncertainty relation
(\ref{gp.01}) introduce the existence of a minimum measurable length.

Thus, from (\ref{gp.01}) we define the deformed algebra
\begin{equation}
\lbrack X_{i},P_{j}]=i\hbar\lbrack\delta_{ij}(1-\beta P^{2})-2\beta P_{i}%
P_{j}]. \label{md1}%
\end{equation}
Consequently, the coordinate representation of the modified momentum operator
is $P_{i}=p_{i}(1-\beta p^{2})$ \cite{Moayedi}, while keeping $X_{i}=x_{i}$
undeformed The pair of variables $x_{i},~p_{i}~$define the canonical
representation satisfying $[x_{i},p_{j}]=i\hbar\delta_{ij}$\thinspace$,$ and
$p^{2}=\gamma^{\mu\nu}p_{\mu}p_{\nu}$. \ 

In the classical level, if $\mathcal{H}=\mathcal{H}\left(  x,p\right)  $ is a
Hamiltonian function, with equations
\begin{equation}
\dot{x}_{i}=\left\{  x_{i},H\right\}  ,~\dot{p}_{i}=\left\{  p_{i},H\right\}
,
\end{equation}
then in the presence of the minimum length due to the deformation algebra
(\ref{md1})~Hamilton's equations are modified as follows \cite{hm1,hm2,hm3}%
\begin{equation}
\dot{x}^{i}=\left(  1-\beta p^{2}\right)  \frac{\partial H}{\partial p_{i}%
}~,~\dot{p}^{i}=\left(  1-\beta p^{2}\right)  \frac{\partial H}{\partial
x_{i}}\text{.}%
\end{equation}

In general, for any modification of the Heisenberg's uncertainty principle in
the form \cite{hm4,hm5,hm6,hm7,hm8}
\begin{equation}
\Delta X_{i}\Delta P_{j}\geqslant\frac{\hbar}{2}[\delta_{ij}(1+\beta f\left(
P\right)  )],
\end{equation}
the corresponding modified Hamilton's equations read%
\begin{equation}
\dot{x}^{i}=\left(  1-\beta f\left(  p\right)  \right)  \frac{\partial
\mathcal{H}}{\partial p_{i}}~,~\dot{p}^{i}=\left(  1-\beta f\left(  p\right)
\right)  \frac{\partial\mathcal{H}}{\partial x_{i}}. \label{md3}%
\end{equation}

A detailed discussion on the effects of the deformed algebra on Newtonian
mechanical systems, and specifically on Kepler's problem, is presented in
\cite{hm2}.

Recently, in \cite{angup}, the Szekeres system was modified in the presence of
the minimum length. The modified Raychaudhuri equation was derived for an
arbitrary function $f\left(  p \right)  $. Specifically, it was found that the
deformed algebra leads to a modified Szekeres system capable of describing
cosmic acceleration. Moreover, the spatial curvature of the spacetime is
affected by the presence of the minimum length and the value of the deformed
parameter $\beta$. In the simplest scenarios, where $f\left(  p \right)  $ is
constant, the modified system admits the integrability properties of the
original system. However, when deriving the modified Szekeres system, the
physical parameters depend on the deformed parameter, which introduces a
varying dark energy model. The Szekeres universes with the cosmological
constant term are also examined.

Working in the same spirit, in the following section we employ the effects of
GUP in the modification of the field equations for the $\Lambda$CDM model in
an isotropic and homogeneous FLRW universe.

\section{$\Lambda$-Cosmology with GUP-Modified Raychaudhuri Equation}

\label{sec3}

The field equations of $\Lambda$CDM model are consisted by the conservation
law for the cold dark matter and the Raychaudhuri equation, that is%
\begin{align}
\dot{\rho}_{m}+3H\rho_{m}  &  =0,\label{fe.01}\\
\dot{H}+\frac{H^{2}}{9}+3\left(  \frac{1}{2}\rho-\Lambda\right)   &  =0
\label{fe.02}%
\end{align}
with constraint the Friedmann equation
\begin{equation}
3H^{2}=\rho_{m}+\Lambda. \label{fe.03}%
\end{equation}

To be consistent with the recent analysis of the Szekeres system within the
framework of GUP \cite{angup}, we follow the same procedure for the function
in (\ref{fe.02}). Indeed, from (\ref{fe.01}) we derive $H = -\frac{1}{3}
\left(  \ln\left(  \rho_{m} \right)  \right)  ^{\cdot}$, and by substituting
this into (\ref{fe.02}), we obtain a second-order ordinary differential
equation for the energy density $\rho_{m}$, that is,%
\begin{equation}
\frac{\ddot{\rho}_{m}}{\rho_{m}}+\Lambda-\frac{1}{2}\rho_{m}+\frac{4}%
{3}\left(  \frac{\dot{\rho}_{m}}{\rho_{m}}\right)  ^{2}=0. \label{fe.04}%
\end{equation}

Equation (\ref{fe.04}) follows from the variation of the Action Integral%
\begin{equation}
S=\int\mathcal{L}\left(  t,\rho_{m},\dot{\rho}_{m}\right)  dt, \label{fe.05}%
\end{equation}
in which $\mathcal{L}\left(  t,\rho_{m},\dot{\rho}_{m}\right)  $ is the
point-like Lagrangian of the field equations defined as%
\begin{equation}
\mathcal{L}\left(  t,\rho_{m},\dot{\rho}_{m}\right)  =\frac{1}{2}\left(
\dot{\rho}_{m}\rho_{m}^{-\frac{4}{3}}\right)  ^{2}+\frac{3}{2}\rho_{m}%
^{\frac{1}{3}}\left(  1+\frac{\Lambda}{\rho_{m}}\right)  . \label{fe.06}%
\end{equation}

We define the momentum $P=\frac{\partial\mathcal{L}}{\partial\dot{\rho}_{m}}$,
that is, $P=\dot{\rho}_{m}\rho_{m}^{-\frac{8}{3}}$, and we introduce the
Hamiltonian function%
\begin{equation}
\mathcal{H}\left(  t,\rho_{m},\dot{\rho}_{m}\right)  =\frac{1}{2}\rho
_{m}^{\frac{8}{3}}P^{2}-\frac{3}{2}\rho_{m}^{\frac{1}{3}}\left(
1+\frac{\Lambda}{\rho_{m}}\right)  , \label{fe.07}%
\end{equation}
where due to the constraint equation (\ref{fe.03}) it follows $\mathcal{H}%
\left(  t,\rho_{m},\dot{\rho}_{m}\right)  \equiv0$.

At this point, it is important to mention that we have not introduced the
lapse function within the FLRW line element. In the presence of the lapse
function, the constraint equation follows from the variation of the point-like
Lagrangian with respect to the lapse. In this case, the point-like Lagrangian
is singular, since there is no momentum term in the Lagrangian related to the
evolution of the lapse function. In this work, for simplicity, we assume
without loss of generality that the lapse function is constant.

Therefore, from the autonomous Hamiltonian function (\ref{fe.07}) we derive
the equations of motion
\begin{equation}
\dot{\rho}_{m}=\left\{  \rho_{m},\mathcal{H}\right\}  ,~\dot{P}=\left\{
P,\mathcal{H}\right\}  , \label{fe.08}%
\end{equation}
where $\left\{  ,\right\}  $ is the Poisson bracket, that is
\begin{equation}
\dot{\rho}_{m}=\rho^{\frac{8}{3}}P,~\dot{P}=3\left(  2\Lambda-\rho\right)
\rho_{m}^{-\frac{5}{3}}+8\rho_{m}^{\frac{5}{3}}P^{2} \label{fe.09}%
\end{equation}

\subsection{GUP-Modified Field Equations}

In the presence of the minimal length the Hamiltonian equations (\ref{fe.08})
are modified as%
\begin{align}
\dot{\rho}_{m}  &  =\left(  1+\beta f\left(  t\right)  \right)  \left\{
\rho_{m},\mathcal{H}\right\}  ,\label{fe.10}\\
\dot{P}  &  =\left(  1+\beta f\left(  t\right)  \right)  \left\{
P,\mathcal{H}\right\}  , \label{fe.11}%
\end{align}
that is, equations (\ref{fe.09}) become%
\begin{align}
\dot{\rho}_{m}  &  =\left(  1+\beta f\left(  t\right)  \right)  \rho^{\frac
{8}{3}}P,\label{fe.12}\\
\dot{P}  &  =\left(  1+\beta f\left(  t\right)  \right)  \left(  3\left(
2\Lambda-\rho\right)  \rho_{m}^{-\frac{5}{3}}+8\rho_{m}^{\frac{5}{3}}%
P^{2}\right)  . \label{fe.14}%
\end{align}

From the latter system we can construct the\ evolution equation for the Hubble
function,
\begin{equation}
\dot{H}+H^{2}=\frac{1}{6}\left(  2\Lambda-\rho_{m}\right)  \left(  1+\beta
f\left(  t\right)  \right)  ^{2}+\frac{H}{\beta}\left(  \ln\left(  1+\beta
f\left(  t\right)  \right)  \right)  ^{\cdot}. \label{fe.15}%
\end{equation}
which can be seen as the GUP-Modified Raychaudhuri equation.

Expanding equation (\ref{fe.15}) around $\beta=0$, it follows%
\begin{equation}
\dot{H}+H^{2}=\frac{1}{6}\left(  2\Lambda-\rho_{m}\right)  \left(  1+2\beta
f\left(  t\right)  \right)  +\beta\dot{f}\left(  t\right)  H+O\left(
\beta^{2}\right)  . \label{fe.16}%
\end{equation}

\section{Asymptotic dynamics}

\label{sec4}

In this section, we study the effects of the GUP within the cosmological
dynamics of the $\Lambda$CDM model. In particular, we perform a detailed
analysis of the phase-space.

We employ the Hubble-normalization approach and we introduce the dimensionless
variables%
\begin{equation}
\Omega_{m}=\frac{\rho_{m}}{3H^{2}},~\Omega_{\Lambda}=\frac{\Lambda}{3H^{2}%
},~\tau=\ln a. \label{fe.17}%
\end{equation}
Let $\Omega_{m},~\Omega_{\Lambda}$ be the new dependent variables and $\tau$
to be the new independent variables, then the constraint equation
(\ref{fe.03}) reads%
\begin{equation}
\Omega_{m}+\Omega_{\Lambda}=1, \label{fe.18}%
\end{equation}
while from the modified GUP-Modified Raychaudhuri equation (\ref{fe.16}) we
derive the evolution equation for the energy density of the cold dark matter%
\begin{equation}
\frac{d\Omega_{m}}{d\tau}=\Omega_{m}\left(  3\left(  \Omega_{m}-1\right)
+\beta\left(  f\left(  t\right)  \left(  2+\beta f\left(  t\right)  \right)
\left(  3\Omega_{m}-2\right)  -2\frac{d}{d\tau}\ln\left(  1+\beta f\left(
t\right)  \right)  \right)  \right)  , \label{fe.19}%
\end{equation}
that is,%
\begin{equation}
\frac{d\Omega_{m}}{d\tau}=3\left(  \Omega_{m}-1\right)  \Omega_{m}+2\Omega
_{m}\beta\left(  f\left(  t\right)  \left(  3\Omega_{m}-2\right)
-\frac{df\left(  t\right)  }{d\tau}\right)  +O\left(  \beta^{2}\right)  .
\label{fe.20}%
\end{equation}

In terms of the new variables, the equation of state parameter for the
effective fluid $w_{eff}=-1-\frac{2}{3}\frac{\dot{H}}{H^{2}}$, reads%
\begin{equation}
w_{eff}=-1+\Omega_{m}+\frac{2}{3}\beta\left(  f\left(  t\right)  \left(
3\Omega_{m}-2\right)  -\frac{df\left(  t\right)  }{d\tau}\right)  +O\left(
\beta^{2}\right)  . \label{fe.21}%
\end{equation}

Consequently, the definition of the function $f\left(  t\right)  $, which
determines the modification of the deformation algebra, affects the entire
dynamics and cosmic evolution.

Recall that for $\beta= 0$, the dynamics of the $\Lambda$CDM model are
recovered. Indeed, the evolution equation for cold dark matter (\ref{fe.20})
with $\beta= 0$ possesses the equilibrium points $\Omega_{m}\left(  P_{1}
\right)  = 1$ and $\Omega_{m}\left(  P_{2} \right)  = 1$, where $P_{1}$
describes the unstable matter-dominated era, and $P_{2}$ is an attractor that
describes the de Sitter universe.

Furthermore, from (\ref{fe.20}) and $\beta= 0$, we calculate $\Omega_{m} =
\frac{1}{1 + c_{1}a^{3}}$. Hence, from (\ref{fe.01}), we find $\rho_{m} =
\rho_{m0}a^{-3}$, and by substituting into (\ref{fe.17}), we obtain the Hubble
function for the $\Lambda$CDM universe%
\begin{equation}
H_{\Lambda CDM}^{2}\left(  a\right)  =H_{0}^{2}\left(  \Omega_{m0}%
a^{-3}+\left(  1-\Omega_{m0}\right)  \right)  \label{fe.22}%
\end{equation}
where $\rho_{m0}=3H_{0}^{2}\Omega_{m0}$ and $\rho_{m0}c_{1}=3H_{0}^{2}%
\Omega_{\Lambda0}$,~where $\Omega_{\Lambda0}=1-\Omega_{m0}$.

The evolution of the equation of state parameter is given by the expression
\begin{equation}
w_{\Lambda CDM}\left(  a\right)  =-1+\frac{\Omega_{m0}a^{-3}}{\Omega
_{m0}a^{-3}+\left(  1-\Omega_{m0}\right)  }, \label{fe.33a}%
\end{equation}

{In the following lines, we introduce a nonlinear function }$f\left(
t\right)  ${. We investigate the corresponding dynamics, and construct
the closed-form expression for the Hubble function. In this study, we assume
that }$f\left(  t\right)  ${ is defined in such a way that the
right-hand side of equation (\ref{fe.20}) is dimensionless. In \cite{angup}
the introduction of a constant parameter }$f\left(  t\right)  ${ in the
Szekeres universe lead to a universe which deviate from that of the undeformed
algebra. The case of the constant function }$f\left(  t\right)  ${ is
examined in Appendix A. Specifically we determined the analytic solution for
the modified GUP where show that there exist a small deviation from the
}$\Lambda${CDM solution. In this study, we focus in a function
}$f\left(  t\right)  \,\ ${which describe a quadratic GUP model
\cite{ali1}. The quadratic GUP model has been widely studied in the literature
and has many applications in gravitational physics. In particular, the
quadratic GUP has been studied to understand the effects of GUP in local
gravity \cite{Vag1,Vag2}, while some cosmological applications can be found in
\cite{ali2,ali3} }

\subsection{Quadratic GUP: $f\left(  t\right)  \simeq P^{2}\rho_{m}^{-\frac
{7}{3}}$}

We consider the quadratic modification of GUP with $f\left(  t\right)
=\frac{1}{9}P^{2}\rho_{m}^{-\frac{7}{3}}$ where $\rho_{m}^{-\frac{7}{3}}$ has
been introduced in order $f\left(  t\right)  $ to be dimensionless. The
deformed function is expressed equivalent as $f\left(  t\right)  =\frac{H^{2}%
}{\rho_{m}}$ or $f\left(  t\right)  =\frac{1}{3\Omega_{m}}$.

By replacing in (\ref{fe.20}) we determine the evolution equation for the
energy density of the cold dark matter%
\begin{equation}
\frac{d\Omega_{m}}{d\tau}=3\Omega_{m}\left(  \Omega_{m}-1\right)  +\left(
4\Omega_{m}-\frac{10}{3}\right)  \beta+O\left(  \beta^{2}\right)  .
\label{fe.26}%
\end{equation}
where the effective equation of state parameter reads%
\begin{equation}
w_{eff}=-1+\Omega_{m}+\frac{2}{3}\left(  2-\frac{5}{3}\frac{1}{\Omega_{m}%
}\right)  \beta+O\left(  \beta^{2}\right)  \label{fe.27}%
\end{equation}
It is important to mention that the $O\left(  \beta^{2}\right)  $ terms in
equation (\ref{fe.26}) introduce the $\Omega_{m}$ parameter in the
denominator. Hence, near the singularity, where $\Omega_{m}\rightarrow0$,
additional nonlinear terms contribute, and our conclusions do not hold for the
global dynamics of the dynamical system. Consequently, in the following, we
assume $\Omega_{m}\gg\beta$.

There exist the unique exact equilibrium point,
\[
B_{1}:\Omega_{m}=1-\frac{2}{3}\beta.
\]
and the family of points%
\[
B_{2}:~\Omega_{m}\simeq\beta\text{.}%
\]
The asymptotic solution near the equilibrium point has the following physical
parameters $\Omega_{\Lambda}\left(  B_{1}\right)  =\frac{2}{3}\beta$%
,~$w_{eff}\left(  B_{1}\right)  =O\left(  \beta^{2}\right)  $. Thus point
$B_{1}$ describes the matter dominated epoch. As far as the stability is
concerned, point $B_{1}$ is a source. On the other hand points $B_{2}$
describe solutions near to the de Sitter spaceime, where $\Omega_{\Lambda
}\left(  B_{2}\right)  \simeq1-\beta$. Point $B_{2}$ is always an attractor.

On the other hand, equation (\ref{fe.26}) leads to the asymptotic solution%
\begin{equation}
\Omega_{m}\left(  a\right)  =\frac{1}{6}\left(  3-4\beta+\sqrt{9+16\beta
}-\frac{2\sqrt{9+16\left(  1+\beta\right)  }}{1+\left(  c_{1}\right)
^{-1}a^{-\sqrt{9+16\beta\left(  1+\beta\right)  }}}\right)  . \label{fe.28}%
\end{equation}
The modified Hubble function in terms of the physical parameters
$H_{0},~\Omega_{m0}$ read%
\begin{equation}
H^{2}\left(  a\right)  =H_{G0}^{2}\frac{3\Omega_{m0}a^{-3}}{\left(
\frac{\sqrt{1+\frac{16}{9}\beta\left(  1+\beta\right)  }\Omega_{m0}a^{-3}%
}{\Omega_{m0}a^{-3}+\left(  1-\Omega_{m0}\right)  a^{3\left(  -1+\sqrt
{1+\frac{16}{9}\beta\left(  1+\beta\right)  }\right)  }}+\frac{1}{2}\left(
1-\sqrt{1+\frac{16}{9}\beta\left(  1+\beta\right)  }\right)  -\frac{2}{3}%
\beta\right)  } \label{fe.29a}%
\end{equation}

Expanding around $\beta\rightarrow0$, it follows%
\begin{equation}
H^{2}\left(  a\right)  =H_{\Lambda CDM}^{2}\left(  a\right)  +\beta
H_{cor}^{B}\left(  a\right)  , \label{fe.30}%
\end{equation}
where
\begin{equation}
H_{cor}^{B}\left(  a\right)  =\frac{2}{3}H_{G0}^{2}\left(  \Omega_{m0}%
a^{-3}+6\left(  1-\Omega_{m0}\right)  +\frac{5\left(  1-\Omega_{m0}\right)
^{2}}{\Omega_{m0}}a^{3}+12\left(  1-\Omega_{m0}\right)  \ln a\right)  .
\label{fe.31}%
\end{equation}

From the above, it is clear that the de Sitter universe is not an attractor,
because when $\Omega_{m0} \rightarrow0$, the correction term dominates.
Therefore, this solution is well-defined for $\Omega_{m0}$ far from zero.

On the other hand, the evolution of the effective equation-of-state parameter
is given by the following function%
\begin{equation}
w_{eff}\left(  a\right)  =w_{\Lambda CDM}\left(  a\right)  +\beta w_{cor}%
^{B}\left(  a\right)  , \label{fe.32}%
\end{equation}
in which the correction is defined as%
\begin{equation}
w_{cor}^{B}=-\frac{2}{9}\frac{\left(  9\Omega_{m0}^{2}\left(  1-\Omega
_{m0}\right)  a^{-3}+14\Omega_{m0}\left(  1-\Omega_{m0}\right)  ^{2}%
+5a^{3}\left(  1-\Omega_{m0}\right)  ^{3}+12\Omega_{m0}^{2}\left(
1-\Omega_{m0}\right)  a^{-3}\ln a\right)  }{\Omega_{m0}\left(  \Omega
_{m0}a^{-3}+\left(  1-\Omega_{m0}\right)  \right)  ^{2}}. \label{fe.33}%
\end{equation}

\begin{figure}[ptbh]
\centering\includegraphics[width=0.5\textwidth]{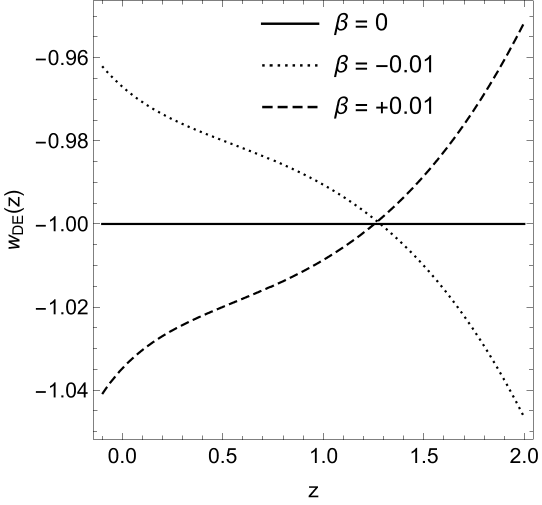}\caption{Qualitative
evolution for the equation of state parameter $w_{DE}\left(  a\right)  $ for
the GUP-modified $\Lambda$CDM model (\ref{fe.29a}) and (\ref{fe.33a}) for
different values of parameter $\beta$. Solid line is for the$~\Lambda$CDM
universe, dotted line is for $\beta=-0.01$ and dashed line is for
$\beta=+0.01$. For the plot we considered $\Omega_{m0}=0.3$. {The
transition point where the }$w_{DE}\left(  a\right)  ${ crosses the
phantom divide line depends on parameter }$\Omega_{m0}.${ For larger
values of }$\Omega_{m0}${ tthe transition point occurs at lower
redshifts, while for smaller values of }$\Omega_{m0}${ it shifts to
higher redshifts.}}%
\label{fig00}%
\end{figure}

We write the Hubble function (\ref{fe.29a}) in the form%
\begin{equation}
H^{2}\left(  a\right)  =H_{0}^{2}\left(  \Omega_{m0}a^{-3}+\Omega
_{DE0}e^{-3\int_{a}^{a_{0}}\frac{\left(  1+w_{DE}\left(  a\right)  \right)
}{a}da}\right)  , \label{fe.33aa}%
\end{equation}
such that to define the $w_{DE}\left(  a\right)  $. The qualitative evolution
of $w_{DE}\left(  a\right)  $ is presented in \ref{fig00}, where we observe
that for $\beta>0$, the phantom epoch is in the present time, while for
$\beta<0$, the phantom epoch was in the past.

\begin{figure}[ptbh]
\centering\includegraphics[width=0.5\textwidth]{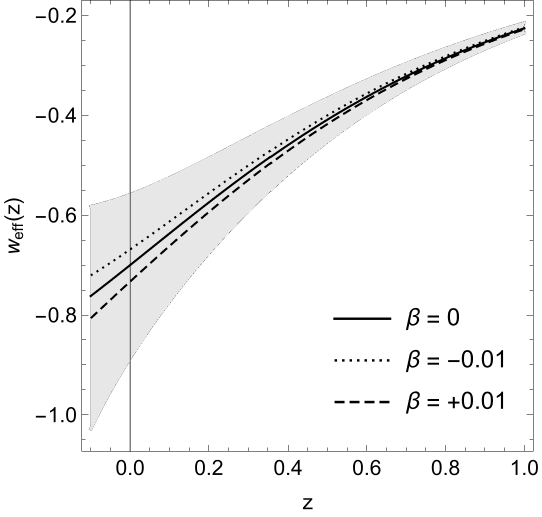}\caption{Qualitative
evolution for the effective equation of state parameter $w_{eff}\left(
a\right)  $ for the GUP-Modified $\Lambda$CDM model (\ref{fe.29a}) for
different values of parameter $\beta$. Gray area is for $\beta\in\left[
-0.05,+0.05\right]  $. Solid line is for the$~\Lambda$CDM universe, dotted
line is for $\beta=-0.01$ and dashed line is for $\beta=+0.01$. For the plot
we considered $\Omega_{m0}=0.3$. }%
\label{fig0}%
\end{figure}

In Fig. \ref{fig0}, we present the qualitative evolution of the effective
equation-of-state parameter for the analytic solution (\ref{fe.29a}), for
different values of the parameter $\beta$. We observe that for negative
$\beta$, the present value $w_{eff}\left(  a\rightarrow1\right)  $ is greater
than that of the $\Lambda$CDM model, which means that the dynamical dark
energy has an equation-of-state parameter whose present value is greater than
$-1$. On the other hand, for positive values of the parameter $\beta$, the
dynamical dark energy has an equation-of-state parameter whose present value
is less than $-1$. \ Furthermore for large values of the redshift, we observe
that $w_{eff}\rightarrow0$, which means the matter dominates.

{We observe that for for }$a<<1${, }$w_{cor}^{B}\left(
a\right)  \rightarrow0${, such that the effective equation of state
parameter }$w_{eff}\left(  a\right)  ${ (\ref{fe.32}) to be near the
the matter epoch as described by point }$B_{1}${. Recall that radiation
has been omitted since we focus in the late-time universe. }

\section{Observational Data Analysis}

\label{sec5}

For our observational analysis, in order to determine the cosmological
parameters, we make use of three background datasets: the Pantheon+ Type Ia
supernova data (SNIa), the Cosmic Chronometers (CC), and the Baryonic Acoustic
Oscillations (BAO) data.

\begin{itemize}
\item The Pantheon+
dataset\footnote{https://github.com/PantheonPlusSH0ES/DataRelease} comprises
1701 light curves of 1550 spectroscopically confirmed supernova events within
the range $10^{-3}<z<2.27$. The data provides the distance modulus $\mu
^{obs}~$at~observed redshifts~$z~$\cite{pan}.~

\item For the Cosmic Chronometers we use 31 direct measurements of the Hubble
parameter across redshifts in the range $0.09\leq z\leq1.965~$ \cite{cc1}.

\item The BAO data are provided from the DESI 2024 collaboration
\cite{des1,des2,des3}, which we shall refer as BAO2024; and of the DESI 2025
collaboration \cite{des4,des5,des6} which we shall refer as BAO2025.
\end{itemize}

The Hubble function that we use for the cosmological analysis is
\begin{equation}
\left(  \frac{H\left(  a\right)  }{H_{0}}\right)  ^{2}=\Omega_{b0}%
a^{-3}+\left(  1-\Omega_{b0}\right)  \left(  \frac{H_{GUP}\left(
a,\Omega_{m0}\right)  }{H_{0}^{GUP}}\right)  ^{2}, \label{fe.34}%
\end{equation}
where $\Omega_{b0}$ is the energy density of the baryons, $H_{GUP}\left(
a,\Omega_{m}\right)  $ is function (\ref{fe.29a}) and $H_{0}^{GUP}$ is a
normalized constant, such that at the present time, $a=1$,~$H_{GUP}\left(
a,\Omega_{m}\right)  =H_{0}^{GUP}\,$, that is, $H\left(  a\right)  =H_{0}$.

To constrain the above parameters with the datasets, we make use of the
Bayesian inference framework COBAYA \cite{cob1,cob2}, with a custom theory and
the PolyChord nested sampler \cite{poly1,poly2}, which provides the Bayesian
evidence an important quantity for statistical comparison between models. For
the energy density of baryons, we adopt the results of the Planck 2018
collaboration \cite{planck}, and assume $\Omega_{b0}\simeq0.0486$. Recall that
the Planck data constraint the $\omega_{b}=\Omega_{b}h^{2}$, with $\omega
_{b}=0.0224$, however, here we consider the previous ansatz.

Hence, we run our model for the free parameters $H_{0},\Omega_{m0},r_{drag}$
and $\beta$. The parameters are constrained as follows: $H_{0}\in\left[
60,75\right]  $, $\Omega_{m0}\in\left[  0.20,0.35\right]  $, $\beta\in\left[
-0.05,0.02\right]  $, and $r_{drag}\in\left[  135,155\right]  $. Although the
parameter $\beta$ is generally considered to be very small, in the analysis of
cosmological data we relax this condition and allow $\left\vert \beta
\right\vert $ to take larger values.

For our statistical analysis, we consider the following sets of data

\begin{itemize}
\item Dataset $\mathbf{D}_{1}:~$SNIa+CC,

\item Dataset$~\mathbf{D}_{2}:~$SNIa+BAO2024,

\item Dataset$~\mathbf{D}_{3}:~$SNIa+BAO2025,

\item Dataset $\mathbf{D}_{4}:~$SNIa+CC+BAO2024,

\item Dataset$~\mathbf{D}_{5}:~$SNIa+CC+BAO2025.
\end{itemize}

Hence, from this analysis we will able to understand the effects of different
sets on the GUP-Modified $\Lambda$CDM Hubble function (\ref{fe.34}).

\subsection{Reference model and information criteria}

The $\Lambda$CDM model is considered as the reference model, where in the
presence of the baryons, the Hubble function reads%
\begin{equation}
\left(  \frac{H_{\Lambda}\left(  a\right)  }{H_{0}}\right)  ^{2}=\Omega
_{b0}a^{-3}+\Omega_{m0}a^{-3}+\left(  1-\Omega_{b0}-\Omega_{m0}\right)
\text{\thinspace}. \label{fe.35}%
\end{equation}

In order to compare the two cosmological theories, that is, the $\Lambda$CDM
model and the GUP-Modified $\Lambda$CDM model we make use of the Akaike
Information Criterion (AIC), as it allows us to compare models with different
dimension $\kappa$ in their parameter space. For a large number of data
points, we compute the AIC from the minimum $\chi^{2}$ value using the
expression \cite{AIC}
\begin{equation}
AIC\simeq-2\ln\mathcal{L}_{\max}+2\kappa.
\end{equation}
The difference of the AIC parameters between two models~$\Delta\left(
AIC\right)  =AIC_{1}-AIC_{2}~$gives information if the models are
statistically indistinguishable. Indeed, if $\left\vert \Delta AIC\right\vert
<2~$the two models are in comparison. Moreover, $\left\vert \Delta
AIC\right\vert >2$ there is a weak evidence than the model with lower
$\chi_{\min}^{2}$ fits the data in a better way. On the other hand, if
$\left\vert \Delta AIC\right\vert >6$ there is strong such evidence, while
when $\left\vert \Delta AIC\right\vert >10$ there exist a clear evidence for
the preference of the model with lover $\chi_{\min}^{2}$ value.

Nevertheless, the Bayesian evidence $\ln\left(  Z \right)  $ can be used to
compare models with different parameter spaces by using Jeffrey's scale
\cite{AIC2}. Let $\Delta\left(  \ln Z \right)  = \ln\frac{Z_{1}}{Z_{2}}$ be
the difference between the Bayesian evidence values for the two models. Then,
if $\Delta\left(  \ln Z \right)  \ll1$, the two models are comparable, and
there is no evidence favoring either model. For $\Delta\left(  \ln Z \right)
< 1$, there is weak evidence in favor of the model with the higher Bayesian
evidence value. When $\Delta\left(  \ln Z \right)  > 1$, the evidence is
moderate; when $\Delta\left(  \ln Z \right)  > 2.5$, the evidence is strong;
and when $\Delta\left(  \ln Z \right)  > 5$, there is very strong evidence
that the model with the higher value, $\ln Z_{2}$, is statistically preferred.

The Akaike's scale and Jeffrey's scale are summarized in\ Table \ref{table0}.%

\begin{table}[tbp] \centering
\caption{Akaike's and Jeffrey's scales.}%
\begin{tabular}
[c]{cccc}\hline\hline
$\mathbf{\Delta}\left(  AIC\right)  $ & \textbf{Akaike's scale} &
$\Delta\left(  \ln Z\right)  $ & \textbf{Jeffrey's scale}\\\hline
$<2$ & Inconclusive & $<<1$ & Inconclusive\\
$2-6$ & Weak (W) evidence & $<1$ & Weak (W) evidence\\
$6-10$ & Strong evidence & $1-2.5$ & Moderate evidence\\
$>10$ & Clear evidence & $>5$ & Clear evidence\\\hline\hline
\end{tabular}
\label{table0}%
\end{table}%

\subsection{Results}

We perform the statistical analysis for five different datasets, and for both
the $\Lambda$CDM model and the GUP-Modified Hubble function (\ref{fe.34}). The
best-fitting parameters with their $1\sigma$ uncertainties, along with the
statistical values of the analysis, are summarized in Table \ref{table1}.
Moreover, in Table \ref{table2}, we present the statistical indicators used to
compare the two cosmological theories.

All datasets provide similar values for the physical parameters $H_{0}$ and
$\Omega_{m0}$, with slightly smaller values for the GUP-Modified model.

The dataset $D_{1}$ is the unique dataset where the $\chi_{\min\Lambda}%
^{2}<\chi_{\min GUP}^{2}$. By using the AIC it follows that there exist a week
evidence for the $\Lambda$CDM. However, the Bayesian evidences for these
models have a difference $\log\frac{\mathbf{Z}_{GUP}}{\boldsymbol{Z}_{\Lambda
}}=0.04$ which means that the two models are statistical equivalent.

By using the data sets $D_{2}$ and $D_{2}$, where we introduce the BAO2024 and
BAO2025 data, it follows that now the $\chi_{\min GUP}^{2}<\chi_{\min\Lambda
}^{2}$. Due to the different number of degrees of freedom, it follows that
AIC$_{GUP}-$AIC$_{\Lambda}=-0.1$ from where we conclude that that the two
models are statistical equivalent. Nevertheless, the Bayesian evidence and
from Jeffrey's scale it follows that $\log\frac{\mathbf{Z}_{GUP}%
}{\boldsymbol{Z}_{\Lambda}}=0.55$ from the BAO2024 data and $\log
\frac{\mathbf{Z}_{GUP}}{\boldsymbol{Z}_{\Lambda}}=0.56$ BAO2025 data, that is,
there exist a week evidence for the strength of the GUP-Modified model. Indeed
there is not any significant difference for the values of the latter between
for the BAO2024 and BAO2025 data.

However, when the CC data are included in datasets $D_{4}$ and $D_{5}$ we
observe that while $\chi_{\min GUP}^{2}$ remains the same, the difference in
the Bayesian evidence becomes greater. Indeed, in comparison with the
$\Lambda$CDM model, the differences in the Bayesian evidence are $0.23$ and
$0.44$ for the BAO2024 and BAO2025 data, respectively. While Jeffrey's scale
indicates weak evidence in favor of the GUP-Modified model, we observe that
this evidence becomes stronger for the $D_{5}$ dataset, where the BAO2025 data
are included.

\begin{figure}[ptbh]
\centering\includegraphics[width=0.7\textwidth]{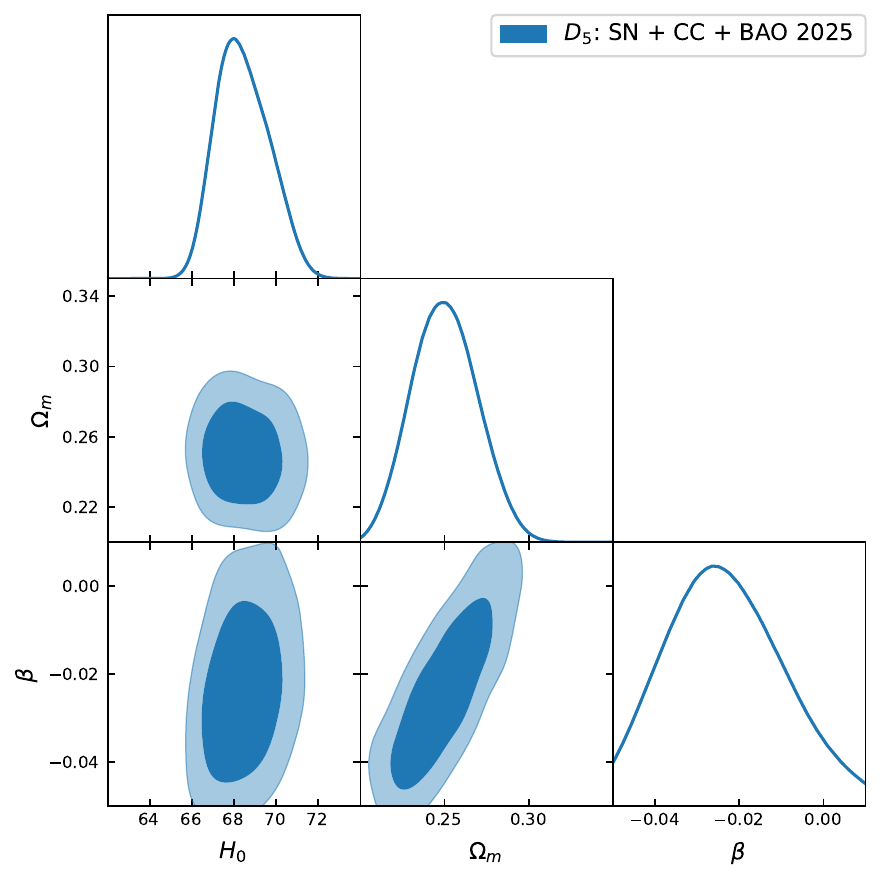}\caption{Confidence
space for the best-fit parameters for the GUP-modified model with Hubble
function (\ref{fe.34}) for the datasets $D_{1}$ (SNIa+CC) and $D_{5}$
(SNIa+CC+BAO2025). }%
\label{fig1}%
\end{figure}

As far as the parameter $\beta$ is concerned, all datasets support that
$\beta$ is negative, with the value $\beta=0$ lying within the $2\sigma$
region. The confidence space for the best-fit parameters for the GUP-Modified
model, for datasets $D_{1}$ and $D_{5}$, are presented in Fig. \ref{fig1}.
Recall that a negative value for the parameter $\beta$ indicates that, at the
present time, the value of the equation-of-state parameter for the dark energy
fluid is greater than $-1$, which is in agreement with previous studies on the
BAO2024 data \cite{db1,db2,db3}. Moreover, as we show in the previous section,
a negative value for parameter $\beta$ indicate the new dynamical dark energy
component can have a phantom behaviour.%

\begin{table}[tbp] \centering
\caption{Obeservation Constraints for the $\Lambda$CDM model and its GUP modification.}%
\begin{tabular}
[c]{ccccccc}\hline\hline
\multicolumn{2}{c}{\textbf{Model/Dataset}} & $\mathbf{D}_{1}$ & $\mathbf{D}%
_{2}$ & $\mathbf{D}_{3}$ & $\mathbf{D}_{4}$ & $\mathbf{D}_{5}$\\\hline
$\Lambda$\textbf{CDM} &  &  &  &  &  & \\
& $\mathbf{H}_{0}$ & $67.9_{-1.7}^{+1.7}$ & $69.3_{-1.4}^{+1.4}$ &
$69.2_{-1.4}^{+1.4}$ & $68.8_{-1.3}^{+1.2}$ & $68.8_{-1.2}^{+1.2}$\\
& $\mathbf{\Omega}_{m0}$ & $0.278_{-0.008}^{+0.020}$ & $0.263_{-0.012}%
^{+0.012}$ & $0.263_{-0.013}^{+0.011}$ & $0.264_{-0.011}^{+0.011}$ &
$0.263_{-0.011}^{+.0.11}$\\
& $\mathbf{\chi}_{\min}^{2}$ & $1419.1$ & $1419.8$ & $1419.9$ & $1434.7$ &
$1434.8$\\
& $\log\mathbf{Z}$ & $-711.07$ & $-712.76$ & $-712.91$ & $-720.40$ &
$-720.47$\\
&  &  &  &  &  & \\
\textbf{GUP} &  &  &  &  &  & \\
& $\mathbf{H}_{0}$ & $67.8_{-1.7}^{+1.7}$ & $69.0_{-1.6}^{+1.6}$ &
$68.9_{-1.8}^{+1.5}$ & $68.5_{-1.5}^{+1.1}$ & $68.4_{-1.5}^{+1.1}$\\
& $\mathbf{\Omega}_{m0}$ & $0.265_{-0.032}^{+0.032}$ & $0.248_{-0.021}%
^{+0.018}$ & $0.249_{-0.021}^{+0.019}$ & $0.250_{-0.020}^{+0.020}$ &
$0.250_{-0.021}^{+0.019}$\\
& $\mathbf{\beta}$ & $-0.018_{-0.015}^{+0.015}$ & $-0.024_{-0.016}^{+0.011}$ &
$-0.024_{-0.015}^{+0.012}$ & $-0.023_{-0.016}^{+0.012}$ & $-0.023_{-0.015}%
^{+0.012}$\\
& $\mathbf{\chi}_{\min}^{2}$ & $1419.4$ & \thinspace$1417.7$ & \thinspace
$1417.8$ & $1433.0$ & $1433.0$\\
& $\log\mathbf{Z}$ & $-711.03$ & $-712.21$ & $-712.36$ & $-720.17$ &
$-720.03$\\\hline\hline
\end{tabular}
\label{table1}%
\end{table}%
%

\begin{table}[tbp] \centering
\caption{Statistical comparison of the two models.}%
\begin{tabular}
[c]{cccccc}\hline\hline
\textbf{Dataset} & $\Delta\left(  \mathbf{\chi}_{\min GUP-\Lambda}^{2}\right)
$ & \textbf{AIC}$_{GUP}-$\textbf{AIC}$_{\Lambda}$ & \textbf{Akaike's scale} &
$\log\frac{\mathbf{Z}_{GUP}}{\boldsymbol{Z}_{\Lambda}}$ & \textbf{Jeffrey's
scale}\\
$\mathbf{D}_{1}$ & $0.3$ & $2.3$ & W evidence for $\Lambda$ & $0.04$ &
Incoclusive\\
$\mathbf{D}_{2}$ & $-2.1$ & $-0.1$ & Incoclusive & $0.55$ & W evidence for
GUP\\
$\mathbf{D}_{3}$ & $-2.1$ & $-0.1$ & Incoclusive & $0.56$ & W evidence for
GUP\\
$\mathbf{D}_{4}$ & $-1.7$ & $+0.3$ & Incoclusive & $0.23$ & W evidence for
GUP\\
$\mathbf{D}_{5}$ & $-1.8$ & $+0.2$ & Incoclusive & $0.44$ & W evidence for
GUP\\\hline\hline
\end{tabular}
\label{table2}%
\end{table}%

\section{Conclusions}

\label{sec6}

Within the framework of GUP, we modified the cosmological field equations for
the $\Lambda$CDM universe by using the deformed algebra that emerges from the
introduction of a minimum length. The field equations for the $\Lambda$CDM
model admit a minisuperspace description; consequently, they can be written as
a point-like Hamiltonian system. We wrote the modified Hamiltonian equations
and derived the cosmological field equations in the original variables. We
found that new dynamical terms are introduced into the Raychaudhuri equation,
related to the definition of GUP. These new terms drive the dynamics such that
the cosmological constant describes a dynamical dark energy model.

We considered the quadratic GUP expressed in normalized variables. As a
result, nonlinear components are introduced into the modified field equations,
which provide a dynamical dark energy model that unifies the dark sector of
the universe. For this model, we derived a closed-form solution for the Hubble
function in the asymptotic limit where the deformation parameter is very small.

We proposed the latter model as a dark energy candidate and considered it as a
GUP-Modified $\Lambda$CDM theory. We used the Pantheon+ SNIa data, the CC, and
the BAO data provided by the DESI 2024 and DESI 2025 collaborations. With
these datasets, we constrained the free parameters of the model and compared
it to the original $\Lambda$CDM. The statistical analysis shows that the
GUP-Modified $\Lambda$CDM theory, when BAO data are included, fits the
observations better than the standard $\Lambda$CDM model. Although the AIC
suggests that both models fit the data similarly, Jeffrey's scale for Bayesian
evidence indicates weak evidence in favor of the GUP-Modified model. This
evidence becomes stronger when the more recent BAO data release is considered.

\begin{figure}[ptbh]
\centering
\par
\includegraphics[width=0.4\textwidth]{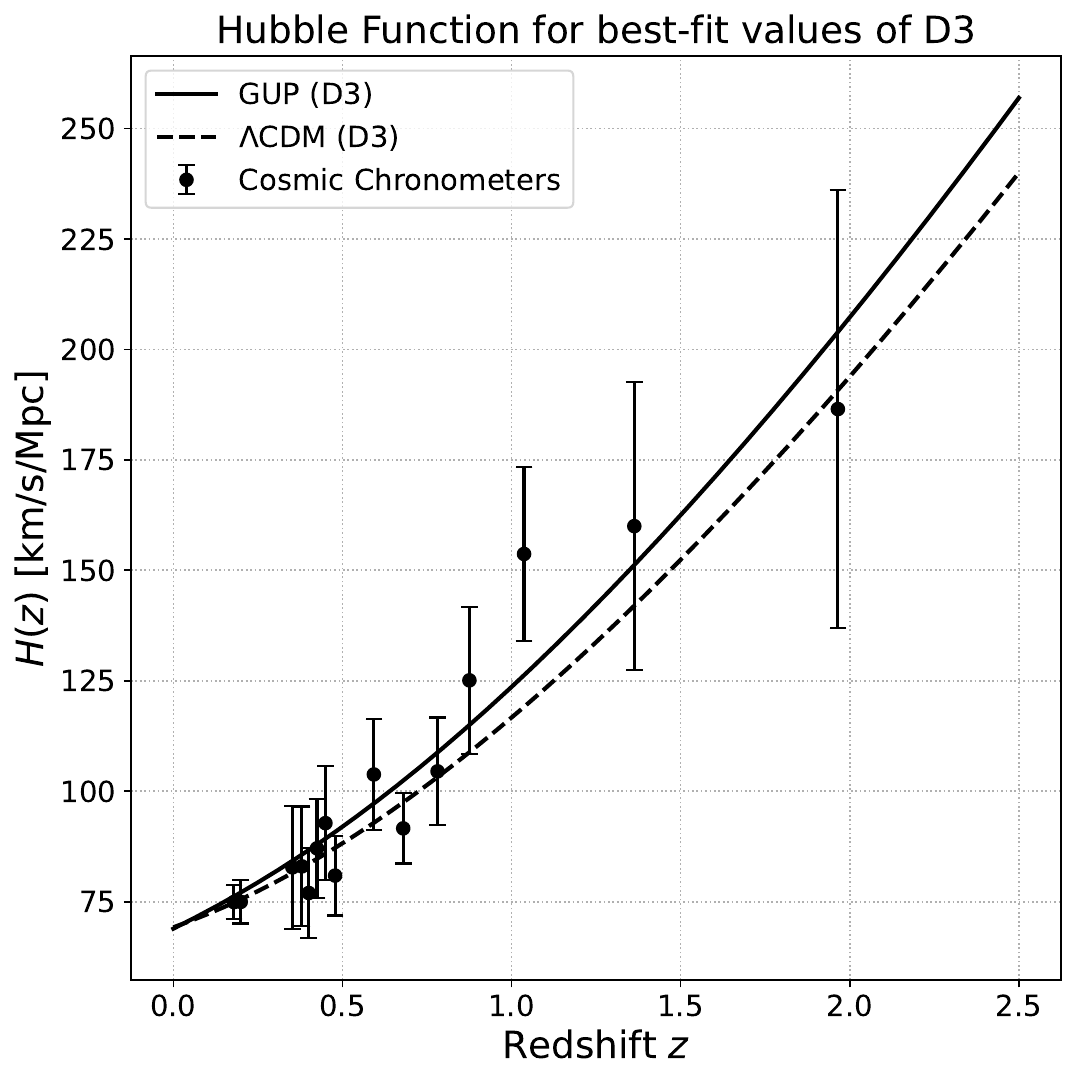}
\includegraphics[width=0.4\textwidth]{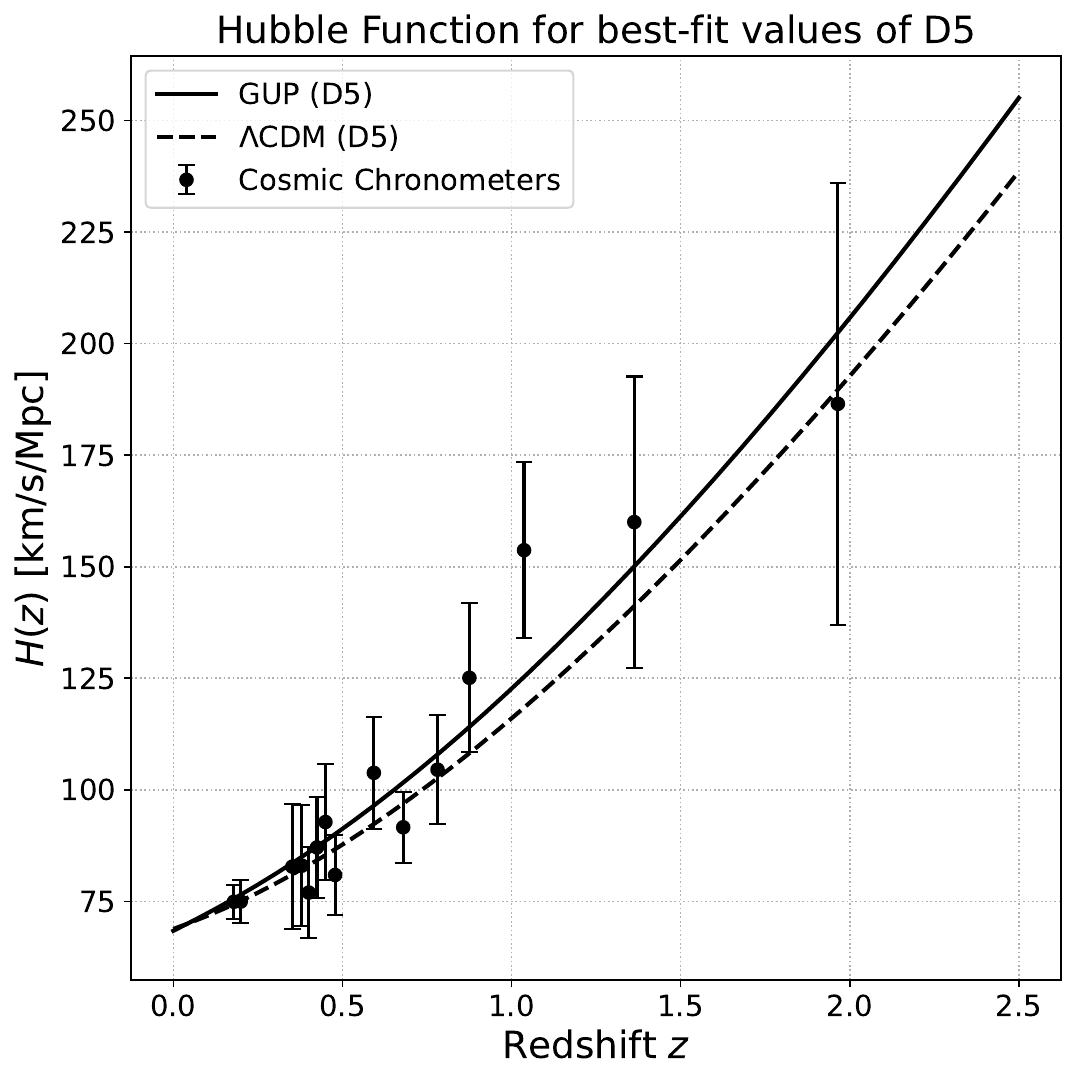} \caption{Theoretical
prediction for the Hubble function $H(z)$ of the GUP and of the $\Lambda$CDM
models. The cosmic chronometers are marked.}%
\label{grid}%
\end{figure}

{In Fig. \ref{grid} we present the theoretical predictions for the
Hubble function using the best-fit parameters of datasets $D_{3}$ and $D_{5}$
within both the $\Lambda$CDM model and the quadratic GUP-modified model.}

This analysis shows that GUP can provide a mechanism for describing dynamical
dark energy models, and specifically, it can be used as a theoretical
framework for introducing a time-varying cosmological constant, starting from
the standard cosmological constant. {Last but not least, we have used
the cosmological observations to constraint the deformation parameter }$\beta
${. We have seen that for this specific deformation algebra,
cosmological data support a negative value for }$\beta${ where the
}$\left\vert \beta\right\vert ${ fits within the bounds for the value
of the deformation parameter for the quadratic GUP determined by other
gravitational tests \cite{gp1,gp2,gp3,gp4,gp5}.}

In a future study, we plan to investigate a more general formulation of the
GUP modification, as well as extend this approach to the Extended Uncertainty
Principle \cite{eup,eup1}. Finally, the cosmological tensions will be
addressed elsewhere \cite{h0t,h0t1,h0t2}.

\begin{acknowledgments}
The authors thanks the support of VRIDT through Resoluci\'{o}n VRIDT No.
096/2022 and Resoluci\'{o}n VRIDT No. 098/2022. Part of this study was
supported by FONDECYT 1240514. The authors thanks Dr. F. Anagnostopoulos for
valuable comments and suggestions, Dr. N. Dimakis and the Universidad de La
Frontera for the hospitality provided when this work was carried out.
\end{acknowledgments}
\appendix

\section{Constant function $f\left(  t\right)  =1$}

We follow \cite{angup} and consider the simplest modification of the
deformation algebra, where $f\left(  t\right)  $ is a constant parameter, that
is, $f\left(  t\right)  =1$. Hence, equation (\ref{fe.20}) reads%
\begin{equation}
\frac{d\Omega_{m}}{d\tau}=3\left(  \Omega_{m}-1\right)  \Omega_{m}+2\Omega
_{m}\beta\left(  \left(  3\Omega_{m}-2\right)  \right)  +O\left(  \beta
^{2}\right)  , \label{fe.23}%
\end{equation}
The equilibrium points points are
\begin{align*}
A_{1}  &  :\Omega_{m}=1-\frac{2}{3}\beta~,\\
A_{2}  &  :\Omega_{m}=0\text{\thinspace}.
\end{align*}

Equilibrium point $A_{1}$ modifies the matter-dominated point $P_{1}$, where
there now exists a small contribution from the cosmological constant in the
universe, such that $\Omega_{\Lambda}\left(  A_{1}\right)  =\frac{2}{3}\beta$
and $w_{eff}\left(  A_{1}\right)  =O\left(  \beta^{2}\right)  $. Indeed, for
$\beta<0$ there exists a phantom behaviour for the dynamical dark energy model.

On the other hand, $A_{2}$ describes a universe dominated by the cosmological
constant; however, now $w_{eff}\left(  A_{2}\right)  =-1-\frac{2}{3}\beta$,
which means that the GUP-Modified model can cross the phantom divide line.

To investigate the stability properties of the points, we study the
eigenvalues of the linearized equation (\ref{fe.23}) around the equilibrium
points. We calculate the eigenvalues $e\left(  A_{1} \right)  = 3 + 4\beta$
and $e\left(  A_{2} \right)  = -3 - 4\beta$. Thus, since $\beta^{2}
\rightarrow0$, point $A_{1}$ is always a source, and $A_{2}$ is always an attractor.

However, the nonlinear differential equation (\ref{fe.23}) can be solved
explicitly, that is,
\begin{equation}
\Omega_{m}\left(  a\right)  =\frac{3+4\beta}{3\left(  1+2\beta\right)
+3c_{1}a^{3+4\beta}}. \label{fe.24}%
\end{equation}
Therefore, from (\ref{fe.01}) and (\ref{fe.17}) \ we calculate%
\begin{equation}
H^{2}\left(  a\right)  =H_{0}^{2}\frac{3\left(  1+2\beta\right)  \Omega
_{m0}+\left(  1-\Omega_{m0}\right)  a^{4\beta}}{3+4\beta}%
\end{equation}

The latter Hubble function expands around~$\beta=0$ as follows%
\begin{equation}
H^{2}\left(  a\right)  =H_{\Lambda CDM}\left(  a\right)  +\beta H_{cor}%
^{A}\left(  a\right)  , \label{fe.25b}%
\end{equation}
in which the correction term is defined as
\begin{equation}
H_{cor}^{A}=\frac{2}{3}H_{0}^{2}\left(  \Omega_{m0}a^{-3}-2\left(
1-\Omega_{m0}\right)  \left(  1+3\ln a\right)  \right)  \text{.}%
\end{equation}

We observe that what has been modified is the cosmological constant term,
which now is no longer a constant. This is in agreement with the additional
terms introduced in the Raychaudhuri equation (\ref{fe.16}) and can be related
to a modified pressure component for the cosmological constant, such that the
equation of state for dark energy differs from $-1$.

\end{document}